\documentclass[10pt,twocolumn,twoside]{IEEEtran}
\IEEEoverridecommandlockouts

\usepackage{cite}
\usepackage{url}
\usepackage{amsmath,amssymb,amsfonts}
\usepackage{algorithmic}
\usepackage{graphicx}
\usepackage{textcomp}
\usepackage{balance}
\usepackage{booktabs}
\usepackage{multirow}
\usepackage{xcolor}
\usepackage[utf8]{inputenc}
\usepackage{tikz}
\usetikzlibrary{calc,arrows.meta, decorations.pathreplacing}
\usepackage{amsthm}
\usepackage{subcaption}

\usepackage{xcolor}
\definecolor{strongblue}{RGB}{31,78,121}
\definecolor{rustorange}{RGB}{191,87,0}

\graphicspath{{../figures/}{figures/}}

\usepackage{array}
\newcolumntype{P}[1]{>{\raggedright\arraybackslash}p{#1}}

\newcounter{propcounter}
\newcounter{corcounter}
\newcounter{theocounter}
\newcounter{defcounter}
\newcounter{asscounter}
\newcommand{\assumption}[1]{%
  \refstepcounter{asscounter}%
  \textbf{Assumption \theasscounter\ (#1).}\ }
\newcommand{\definition}[1]{%
  \refstepcounter{defcounter}%
  \textbf{Definition \thedefcounter\ (#1).}\ }
\newcommand{\proposition}[1]{%
  \refstepcounter{propcounter}%
  \textbf{Proposition \thepropcounter\ (#1).}\ }
\newcommand{\corolary}[1]{%
  \refstepcounter{corcounter}%
  \textbf{Corollary \thecorcounter\ (#1).}\ }
\newcommand{\theorem}[1]{%
  \refstepcounter{theocounter}%
  \textbf{Theorem \thetheocounter\ (#1).}\ }

\def\BibTeX{{\rm B\kern-.05em{\sc i\kern-.025em b}\kern-.08em
    T\kern-.1667em\lower.7ex\hbox{E}\kern-.125emX}}
\usepackage{lipsum}

\title{\LARGE \bf 
Structural Misalignment in Financial Transmission Rights
}

\author{
Erich Trieschman$^\dagger$, Saurabh Amin$^\dagger$
\vspace{-0.5cm}
 \thanks{ 
$^\dagger$First and second authors are with the Department of Civil and Environmental Engineering, 
Massachusetts Institute of Technology, Cambridge, Massachusetts, USA. Email: {\tt \{etriesch, amins\}@mit.edu}}
}

\begin{document}
\begingroup
\allowdisplaybreaks

\maketitle

\begin{abstract}
Financial Transmission Rights (FTRs) enable electricity market participants to hedge congestion risk in Day Ahead Market (DAM) operations, but for the market to be solvent, Independent System Operators (ISOs) must ensure that FTR payouts do not exceed the collected DAM merchandising surplus that funds them. We show that FTR underfunding (or conversely, hedging efficiency) can arise structurally from misalignment between the network models used in the FTR auction and the DAM, independent of bidding behavior.

We develop a geometric framework in which both DAM merchandising surplus and the maximum supportable FTR payout are expressed as support functions of network-feasible injection polytopes. The resulting dual representation assigns nonnegative weights to transmission element–contingency constraints, enabling constraint-level attribution of model misalignment.

Using this framework, we derive sharp implications for canonical FTR network modeling choices like uniform transmission element derates, and for structural sources of underfunding like unplanned DAM outages. We further show that multi-interval FTR products impose an intrinsic hedging inefficiency when DAM shadow prices vary over time, even under perfect model alignment.

These results provide ISOs with rigorous tools to diagnose underfunding and quantify the efficiency cost of conservative FTR network modeling choices.
\end{abstract}


\section{Introduction}
The Financial Transmission Right (FTR) auction allocates congestion-related price risk across wholesale electricity market participants ahead of real-time operations. An FTR pays the Day Ahead Market (DAM) locational marginal price (LMP) difference between a source node and a sink node over a contract period \cite{Hogan2002FTRFormulations}. These payments are funded by DAM merchandising surplus (realized congestion rents) collected by the Independent System Operator (ISO) when transmission elements operate at their limits (bind) \cite{Hogan2002FTRFormulations}.

FTRs provide a congestion risk hedge for physical participants (e.g., a generator hedged at a hub still faces congestion risk between its bus and the hub), transfer congestion risk to financial participants, and allocate uncertain future DAM merchandising surplus through a forward market \cite{Oren1995, Hogan2002FTRFormulations}.

To manage revenue adequacy, ISOs impose a simultaneous feasibility test (SFT).
The SFT uses a network model to link FTR positions to physical constraints and enables system-wide trading across arbitrary node pairs without separate bilateral instruments \cite{Hogan2002FTRFormulations, Oren1995}.

If the FTR network model matched the realized DAM network model, the auction could (in principle) pass through DAM merchandising surplus fully to FTR holders while remaining solvent. 
In practice, several factors prevent such alignment.
The FTR auction clears using a single network model across the contract period, fixed months or years in advance of realized DAM operations. Heterogeneity in realized DAM network models across the contract period and unplanned DAM outages or reconfigurations can both drive misalignment. 
These challenges are amplified by modern grid dynamics: deep interconnection queues lead to frequent topology changes \cite{lbnl2025QueuedUp}, while renewable integration and large data-center loads are increasing operational uncertainty \cite{ERCOTGIS, NERC2025LargeLoadFERC}.
Misalignment creates two distinct failures. 
If the FTR network model is more permissive than realized DAM feasibility, awarded payouts can exceed collected DAM merchandising surplus resulting in underfunding;
if the FTR network model is more restrictive, participants cannot fully hedge available DAM merchandising surplus resulting in hedging inefficiency \cite{spp2024som, potomac2024ercot}. ISOs therefore face a fundamental design tradeoff between aggressive assumptions exposing underfunding risk and conservative assumptions causing hedging inefficiency.

Recent market outcomes illustrate both sides of this tradeoff. In the Southwest Power Pool (SPP), 2024 Transmission Congestion Right (TCR) funding fell below the stated 90-100\% target range (83\%), and the market monitor attributes much of the shortfall to network model outage gaps between the TCR auction and the DAM \cite{spp2024som}. Prior to August 2024 when SPP tightened outage coordination requirements, transmission owners were only required to submit planned outages above a certain duration and size 14 days in advance, well after the corresponding TCR auction had taken place \cite{spp2024som}. As a result, roughly 75\% of outages were excluded from the auction network models due to lead-time alone \cite{spp2024som}.
ERCOT reports full monthly CRR funding throughout 2024 \cite{potomac2024ercot}, supported in part by conservative transfer capability assumptions (including a uniform derating schedule) and balancing mechanisms \cite{potomac2024ercot}; nevertheless, its market monitor still identifies unforeseen outages and transmission reductions between the auction model and the DAM as the primary source of underfunding risk \cite{potomac2024ercot}. 

These observations motivate fundamental questions: 
How can the efficiency cost of conservative FTR network modeling choices be quantified?
In an underfunding event, can the ISO systematically attribute the shortfall to specific constraints in the DAM and FTR network models?

We treat alignment as a geometric comparison between two network-feasible injection polytopes: one induced by the realized DAM network model and one induced by the FTR network model.
Congestion rents in the DAM arise endogenously from welfare maximization under network constraints \cite{schweppe1988, Hogan1992}. 
We show that both DAM merchandising surplus and the maximum supportable FTR payout can be expressed as the value of the same linear congestion-rent objective evaluated over their respective feasible sets of network injections.
Alignment therefore reduces to a comparison of these two optima: underfunding arises when the FTR-based maximum exceeds the DAM-based maximum, while hedging inefficiency arises when it falls short.
This formulation isolates structural effects of network modeling from bidding behavior, but also highlights a central challenge: realistic networks yield high-dimensional polytopes with substantial dual degeneracy, requiring principled constraint-level attribution.

We develop a geometric and dual-based framework for analyzing structural alignment between the FTR auction and the DAM, with the following contributions:

\emph{Support-function formulation.} 
We formalize the alignment problem as a support function comparison over two network-feasible polytopes, yielding a unified geometric representation of revenue adequacy and hedging inefficiency.
    
\emph{Dual attribution with a shared feasible set.}
We derive LP dual representations in which DAM and FTR model alignment problems share a common dual-feasible set and differ only through their objective cost vectors, enabling principled constraint-level comparisons of alternative network models.

\emph{Robust constraint classification.}
We characterize robust multiplier ranges under dual degeneracy and classify each constraint as \emph{definitely binding}, \emph{degenerately binding}, or \emph{definitely slack} in a given DAM realization, enabling constraint-level diagnostics even when dual optima are non-unique.

\emph{Implications for FTR design.}
We show that: (i) uniform constraint limit derates produce inefficiency proportional to the derate factor; (ii) adding contingencies to the FTR network model can induce hedging inefficiency but does not permit underfunding risk; and (iii) omitting contingencies can permit underfunding but does not induce hedging inefficiency.

\emph{Inefficiency from multi-interval products.}
We show that temporal aggregation across DAM intervals cannot increase underfunding risk but can strictly reduce hedging efficiency when congestion patterns vary over time, even under perfect structural alignment.

These contributions enable ISOs to quantify the efficiency cost of conservative FTR network modeling choices and to systematically attribute observed underfunding to specific contingencies or constraints in the DAM and FTR network models using dual information derived from realized DAM outcomes. The remainder of the paper is organized as follows: Section~\ref{sec:network} formalizes the network models and market-clearing problems. Sections~\ref{sec:support}--\ref{sec:alignment} develop the theoretical framework and main results, and Section~\ref{sec:conclusion} discusses extensions and concludes. Proofs for all stated propositions and theorems are provided in Appendix~\ref{app:proofs}.

\section{\label{sec:network}Network Models and Market Clearing}

This section establishes notation, defines network-feasible injection sets, and characterizes market-clearing outcomes underlying both the DAM and the FTR auction. Throughout, we adopt standard DC optimal power flow (DCOPF) conventions, under which network feasibility is expressed through linear constraints on nodal injections \cite{Hogan2002FTRFormulations}, and focus on a single DAM interval unless otherwise noted.

\subsection{Network Topology Representation}

We fix the following sets common to both markets: $\mathcal{N}$: set of nodes (buses), with $|\mathcal{N}| = n$; $\mathcal{L}$: set of monitored transmission elements (lines and transformers), with $|\mathcal{L}| = \ell$;  $\mathcal{C} = \{c_1, c_2, \ldots, c_m\}$: universal set of contingencies, with $|\mathcal{C}| = m$, indexed in a fixed order. Each $c \in \mathcal{C}$ indexes a specific network topology, such as the base case or an $N\!-\!1$ outage of a transmission element.

For each contingency $c \in \mathcal{C}$, the Power Transfer Distribution Factor (PTDF) matrix, $K_c\in \mathbb R^{\ell\times n}$, denotes a fixed linear mapping from nodal injections, $q\in\mathbb R^n$ (restricted to the balanced subspace), to line flows, $K_cq\in\mathbb R^\ell$; all network topology assumptions required to construct $K_c$ are subsumed into the matrix itself \cite{Hogan2002FTRFormulations}.

For each contingency $c \in \mathcal{C}$ and transmission element $\ell \in \mathcal{L}$, let $\underline{b}_{c,\ell}$ and $\overline{b}_{c,\ell}$ denote the lower and upper flow limits, respectively. These limits may reflect thermal, voltage, or stability constraints and may vary across contingencies to capture emergency ratings \cite{Hogan2002FTRFormulations}.

To unify notation across contingencies, we stack PTDF matrices and line limit vectors according to the fixed ordering of $\mathcal C$. Define
\begin{align*}
    K:= \begin{bmatrix}
        K_{c_1}\\\vdots\\K_{c_m}
    \end{bmatrix},
    \quad
    \underline b := \begin{bmatrix}
        \underline b_{c_1}\\ \vdots \\ \underline b_{c_m}
    \end{bmatrix},
    \quad
    \overline b := \begin{bmatrix}
        \overline b_{c_1}\\ \vdots \\ \overline b_{c_m}
    \end{bmatrix},
\end{align*}
with $K \in \mathbb{R}^{m\ell \times n}$ and $\underline{b}, \overline{b} \in \mathbb{R}^{m\ell}$.

\assumption{Common Network Representation}\label{ass:common_network}
The DAM and FTR markets share a common linear network representation: identical sets $\mathcal{N}$, $\mathcal{L}$, $\mathcal{C}$, and PTDF matrices $\{K_c\}_{c \in \mathcal{C}}$. 
Differences between the DAM and FTR models arise exclusively through: (i) \emph{contingency coverage} -- the subset of $\mathcal{C}$ enforced in each market; (ii) \emph{transmission element limits} -- the values of $(\underline{b}_{c}, \overline{b}_{c})$ for enforced contingencies.

\subsection{Market-Specific Network Models}

The DAM enforces a contingency set $\mathcal{C}^{\mathrm{DAM}} \subseteq \mathcal{C}$, and the FTR auction enforces a contingency set $\mathcal{C}^{\mathrm{FTR}} \subseteq \mathcal{C}$. Associated transmission element lower and upper flow limits are
\begin{align*}
    \{(\underline{f}_c, \overline{f}_c)\}_{c \in \mathcal{C}^{\mathrm{DAM}}},\quad 
    \{(\underline{g}_c, \overline{g}_c)\}_{c \in \mathcal{C}^{\mathrm{FTR}}},
\end{align*}
where $\underline{f}_c, \overline{f}_c, \underline{g}_c, \overline{g}_c \in \mathbb{R}^\ell$.

To compare markets with different contingency coverage, we extend each market's limits to the universal contingency set $\mathcal{C}$ by assigning infinite bounds to non-enforced contingencies. 
Stacking limits according to the same fixed ordering as above yields extended transmission element-contingency limits
\begin{align*}
    \underline{f}, \overline{f}, \underline{g}, \overline{g} \in (\mathbb{R} \cup \{-\infty, +\infty\})^{m\ell}.
\end{align*} 

When a limit equals $\pm \infty$, the corresponding inequality constraint is interpreted as absent. Accordingly, constraints are indexed only over finite bounds via the active index sets
\begin{equation*}
\mathcal{J}^+(\overline{b}) := \{ i : \overline{b}_i < +\infty \}, \quad
\mathcal{J}^-(\underline{b}) := \{ i : \underline{b}_i > -\infty \}.
\end{equation*}

\subsection{Network-Feasible Injection Sets}

\definition{Network-Feasible Injection Polytopes}\label{def:network_polytope}
For extended limit vectors $(\underline{b}, \overline{b})$ satisfying $\underline{b} \preceq \overline{b}$, define
\begin{equation*}
\mathcal{Q}_{\mathrm{net}}(\underline{b}, \overline{b})
:= \left\{ q \in \mathbb{R}^n \,\middle|\,
\begin{aligned}
&(Kq)_i \ge \underline{b}_i && \forall i \in \mathcal{J}^-(\underline{b}),\\
&(Kq)_i \le \overline{b}_i && \forall i \in \mathcal{J}^+(\overline{b}),\\
&\mathbf{1}^\top q = 0
\end{aligned}
\right\}.
\end{equation*}

These constraints enforce network physics (via $K$), active transmission element limits under enforced contingencies, and power balance. 
By Assumption~\ref{ass:common_network} the market-specific network-feasible sets are 
\begin{equation*}
\mathcal{Q}^{\mathrm{DAM}}_{\mathrm{net}} := \mathcal{Q}_{\mathrm{net}}(\underline{f}, \overline{f}),
\qquad
\mathcal{Q}^{\mathrm{FTR}}_{\mathrm{net}} := \mathcal{Q}_{\mathrm{net}}(\underline{g}, \overline{g}).
\end{equation*}

\assumption{Non-emptiness}\label{ass:nonempty}  
We assume $\mathbf{0} \in \mathcal{Q}^{\mathrm{DAM}}_{\mathrm{net}}$ and $\mathbf{0} \in \mathcal{Q}^{\mathrm{FTR}}_{\mathrm{net}}$.
Hence both polytopes are nonempty, ensuring the maxima considered later are attained.
A sufficient condition is $\underline f_c\preceq\mathbf0\preceq\overline f_c$ for all enforced contingencies since zero injections induce zero flows. 
In practice, this condition is always satisfied as transmission limits are designed to allow zero flows.

In the DAM, each unit $\omega\in\mathcal U$ submits quantity bounds $[\underline q^u_\omega,\overline q^u_\omega]$ and price $p^u_\omega$, mapped to nodal injections by $M^u_\omega\in\{-1,0,1\}^{n}$ with one nonzero entry.
In the FTR auction, bid $\beta\in\mathcal B$ specifies a source–sink pair with quantity bounds $[\underline q^b_\beta,\overline q^b_\beta]$ and price $p^b_\beta$, mapped by $M^b_\beta\in\{-1,0,1\}^{n}$ with one $+1$ and one $-1$ entry.
DAM prices represent marginal costs (negative for demand), yielding cost minimization; FTR prices represent willingness to pay, yielding value maximization.

\definition{Full Feasible Sets}
Combining network feasibility with a matrix representation of bid availability, the full feasible sets are
\begin{align*}
    \mathcal{Q}^{\mathrm{DAM}} &:= \left\{ 
        q \in \mathbb{R}^n \,\middle|\, q = M^u q^u,\; \underline{q}^u \preceq q^u \preceq \overline{q}^u,\;
        q \in \mathcal{Q}^{\mathrm{DAM}}_{\mathrm{net}}
    \right\},\\
    \mathcal{Q}^{\mathrm{FTR}} &:= \left\{ 
        q \in \mathbb{R}^n \,\middle|\, q = M^b q^b,\; \underline{q}^b \preceq q^b \preceq \overline{q}^b,\;
        q \in \mathcal{Q}^{\mathrm{FTR}}_{\mathrm{net}}
    \right\}.
\end{align*}

We abstract from market features that restrict feasible injections but do not alter network feasibility, such as unit commitment, ramping, and ancillary services in the DAM, and options in the FTR auction. Our focus is on the network constraints that govern structural revenue adequacy and hedging capacity.

\subsection{Market Clearing and Shadow Prices}
The DAM minimizes total bid-in cost subject to its full feasibility set,
\begin{align*}
    \min_{q^u \in \mathbb{R}^{|\mathcal{U}|}} \; (p^u)^\top q^u \qquad
    \text{subject to} \quad q := M^u q^u \in \mathcal{Q}^{\mathrm{DAM}}.
\end{align*}
Let $q^{\text{DAM}*}$ denote the resulting optimal nodal injections.

Let $y^{+*}, y^{-*} \in\mathbb R_+^{m\ell}$ denote the dual multipliers associated with upper and lower bounds on transmission element-contingency constraints respectively, and let $s^*$ denote the dual multiplier on power balance. Define $y^*:= y^{+*} - y^{-*}$, the DAM shadow-price vector. The LMP vector is
\begin{align*}
    \lambda := s^*\mathbf 1 - K^\top y^* \in \mathbb R^n,
\end{align*}
which decomposes LMPs into an energy component $s^*$ and a congestion adjustment $K^\top y^*$ \cite{schweppe1988, Hogan1992}.

\definition{DAM Merchandising Surplus}
The congestion rents collected in the DAM by the ISO are precisely
\begin{align*}
    \mathrm{MS}^{\mathrm{DAM}}(y^*) := -\lambda^\top q^{(\mathrm{DAM})*} = y^{*\top}K q^{(\mathrm{DAM})*},
\end{align*}
where the equality follows from power balance \cite{Hogan2002FTRFormulations}.

The FTR auction maximizes total bid value subject to its full feasibility set
\begin{align*}
    \max_{q^b \in \mathbb{R}^{|\mathcal{B}|}} \; (p^b)^\top q^b \qquad
    \text{subject to}\quad q := M^b q^b \in \mathcal{Q}^{\mathrm{FTR}} .
\end{align*}
Let $q^{\text{FTR}*}$ denote the resulting optimal nodal injections.

\definition{FTR Payout}
For a realized DAM shadow-price vector $y^*$ and LMPs $\lambda$, the payout of DAM merchandising surplus to FTR positions is
\begin{equation*}
\mathrm{PO}^{\mathrm{FTR}}(y^*) := -\lambda^\top q^{(\mathrm{FTR})*}
= y^{*\top} K q^{(\mathrm{FTR})*},
\end{equation*}
again where the equality follows from power balance.

\definition{Maximum FTR Payout}\label{def:max_ftr_po}
For a realized DAM shadow-price vector $y^*$, the maximum supportable FTR payout is
\begin{align*}
    \mathrm{PO}^{\mathrm{FTR}}_{\max} (y^*) := 
    \max_{q \in \mathcal Q^{\mathrm{FTR}}_{\mathrm{net}}} y^{*\top} K q.
\end{align*}
By construction, $\mathrm{PO}^{\mathrm{FTR}}(y^*) \leq \mathrm{PO}^{\mathrm{FTR}}_{\max} (y^*)$ for any $q^{(\mathrm{FTR})*}$.

\section{\label{sec:support}Structural Alignment via Support Functions}
We show that both DAM merchandising surplus and the maximum supportable FTR payout coincide with maximizations of a linear functional over the network-feasible polytopes defined in Section~\ref{sec:network} (i.e., support-function evaluations \cite{Boyd2004ConvexOptimization}).
This identification unifies revenue adequacy and hedging efficiency within a single geometric framework and enables the dual-based attribution developed in Section~\ref{sec:dual}.

\definition{Network Support Function}\label{def:net_support}  
For extended limit vectors $(\underline{b}, \overline{b})$ and arbitrary price vector $y$ (not necessarily a DAM shadow-price vector), define
\begin{equation*}
    \phi(\underline{b}, \overline{b}; y)
    := \max_{q \in \mathcal{Q}_{\mathrm{net}}(\underline{b}, \overline{b})} y^\top K q .
\end{equation*}
Under Assumption~\ref{ass:nonempty}, the maximum is attained.
Geometrically, $\phi(\underline{b}, \overline{b}; y)$ is the signed distance from the origin to the supporting hyperplane of $\mathcal{Q}_{\mathrm{net}}(\underline{b}, \overline{b})$ in the direction $K^\top y$.

\definition{Alignment Gap}\label{def:alignment}   
For arbitrary price vector $y$, define
\begin{equation*}
    \Delta(y) := \phi(\underline g,\overline g;y) - \phi(\underline f,\overline f;y).
\end{equation*}

Thus $\Delta(y)$ records whether $\mathcal Q^{\mathrm{FTR}}_{\mathrm{net}}$ extends further ($\Delta(y)>0$) or lies within ($\Delta(y)<0$) $\mathcal Q^{\mathrm{DAM}}_{\mathrm{net}}$ in direction $K^\top y$. 
Figure~\ref{fig:geometric_interpretation} illustrates both cases.

\begin{figure}[h]
\centering
    \footnotesize
    \begin{tikzpicture}[scale=0.8, line join=round, line cap=round]
    \def\sx{6}
    \def\cx{0.0}
    \def\cy{0.0}
    \def\rot{0}

    \def\Rout{3}
    \def\Rin{1.8}

    \coordinate (i1) at ({\cx + \Rout*cos(\rot)},        {\cy + \Rout*sin(\rot)});
    \coordinate (i2) at ({\cx + \Rout*cos(\rot + 60)},   {\cy + \Rout*sin(\rot + 60)});
    \coordinate (i3) at ({\cx + \Rout*cos(\rot + 120)},  {\cy + \Rout*sin(\rot + 120)});
    \coordinate (i4) at ({\cx + \Rout*cos(\rot + 180)},  {\cy + \Rout*sin(\rot + 180)});
    \coordinate (i5) at ({\cx + \Rout*cos(\rot + 240)},  {\cy + \Rout*sin(\rot + 240)});
    \coordinate (i6) at ({\cx + \Rout*cos(\rot + 300)},  {\cy + \Rout*sin(\rot + 300)});

    \coordinate (j1) at ({\cx + \Rin*cos(\rot)},        {\cy + \Rin*sin(\rot)});
    \coordinate (j2) at ({\cx + \Rin*cos(\rot + 60)},   {\cy + \Rin*sin(\rot + 60)});
    \coordinate (j3) at ({\cx + \Rin*cos(\rot + 120)},  {\cy + \Rin*sin(\rot + 120)});
    \coordinate (j4) at ({\cx + \Rin*cos(\rot + 180)},  {\cy + \Rin*sin(\rot + 180)});
    \coordinate (j5) at ({\cx + \Rin*cos(\rot + 240)},  {\cy + \Rin*sin(\rot + 240)});
    \coordinate (j6) at ({\cx + \Rin*cos(\rot + 300)},  {\cy + \Rin*sin(\rot + 300)});

    \coordinate (j1b) at ($(j1) + (-\Rin + 1.4*\Rout, 0)$);
    \coordinate (j2b) at ($(j2) + (-\Rin + 1.4*\Rout, 0)$);
    \coordinate (j6b) at ($(j6) + (-\Rin + 1.4*\Rout, 0)$);

    \coordinate (O) at ({\cx},{\cy});
    \fill[gray!35, opacity=0.45]
    (i1) -- (i2) -- (i3) -- (i4) -- (i5) -- (i6) -- cycle;
    \draw[thick]
    (i1) -- (i2) -- (i3) -- (i4) -- (i5) -- (i6) -- cycle;

    \fill[white, opacity=0.70]
    (j1b) -- (j2b) -- (j3) -- (j4) -- (j5) -- (j6b) -- cycle;
    \draw[thick]
    (j1b) -- (j2b) -- (j3) -- (j4) -- (j5) -- (j6b) -- cycle;

    \fill[black] (O) circle (2pt);
    \node[below right] at (O) {$O$};

    \node[above right] at ($(i5) + (+0.00*\Rin, -0.0*\Rin)$) {$\mathcal Q_{\text{net}}^{\text{DAM}}$};
    \node[above right] at ($(j5) + (+0.00*\Rin, -0.0*\Rin)$) {$\mathcal Q_{\text{net}}^{\text{FTR}}$};

    \coordinate (n0) at ($(i3)$);
    \coordinate (n1) at ($(i3) + 0.15*(-1.0, 3)$);
    \coordinate (m0) at ($(j3)$);
    \coordinate (m1) at ($(j3) + 0.15*(-1.0, 3)$);

    \draw[strongblue, thick, -{Stealth[inset=0pt, length=10pt, angle=90:3pt]}] (n0) -- (n1);
    \draw[strongblue, thick, -{Stealth[inset=0pt, length=10pt, angle=90:3pt]}] (m0) -- (m1);

    \path let \p1 = ($(n1)-(n0)$) in
    coordinate (h0) at ($(n0) + (-\sx*\y1, \sx*\x1)$)
    coordinate (h1) at ($(n0) + (\sx*\y1,-\sx*\x1)$);
    \path let \p1 = ($(m1)-(m0)$) in
    coordinate (g0) at ($(m0) + (-\sx*\y1, \sx*\x1)$)
    coordinate (g1) at ($(m0) + (\sx*\y1,-\sx*\x1)$);

    \draw[draw=strongblue, thick, dotted] 
        (h0) -- (h1)
        node[pos=0.1, sloped, above, fill=white, fill opacity=0.75] {$\mathbf{\textcolor{strongblue}{y}}^\top K q = \phi(\underline f, \overline f;\mathbf{\textcolor{strongblue}{y}})$};
    \draw[draw=strongblue, thick, dashed] 
        (g0) -- (g1)
        node[pos=0.1, sloped, above, fill=white, fill opacity=0.75] {$\mathbf{\textcolor{strongblue}{y}}^\top K q = \phi(\underline g, \overline g;\mathbf{\textcolor{strongblue}{y}})$};

    \draw[draw=strongblue, thick, decorate, decoration={brace, amplitude=15pt, raise=0pt}]
        (i3) -- (j3)
        node[midway, yshift=15pt, sloped, above, fill=white, fill opacity=0.75] {$\Delta(\mathbf{\textcolor{strongblue}{y}})$};

    \coordinate (n0) at ($(i1)$);
    \coordinate (n1) at ($(i1) + 0.5*(1, 0)$);
    \coordinate (m0) at ($(j1b)$);
    \coordinate (m1) at ($(j1b) + 0.5*(1,0)$);

    \draw[rustorange, thick, -{Stealth[inset=0pt, length=10pt, angle=90:3pt]}] (n0) -- (n1);
    \draw[rustorange, thick, -{Stealth[inset=0pt, length=10pt, angle=90:3pt]}] (m0) -- (m1);

    \path let \p1 = ($(n1)-(n0)$) in
    coordinate (h0) at ($(n0) + (-\sx*\y1, \sx*\x1)$)
    coordinate (h1) at ($(n0) + (\sx*\y1,-\sx*\x1)$);
    \path let \p1 = ($(m1)-(m0)$) in
    coordinate (g0) at ($(m0) + (-\sx*\y1, \sx*\x1)$)
    coordinate (g1) at ($(m0) + (\sx*\y1,-\sx*\x1)$);

    \draw[draw=rustorange, thick, dotted] 
        (h0) -- (h1)
        node[pos=0.85, sloped, below, fill=white, fill opacity=0.75] {$\mathbf{\textcolor{rustorange}{y}}^\top K q = \phi(\underline f, \overline f;\mathbf{\textcolor{rustorange}{y}})$};
    \draw[draw=rustorange, thick, dashed] 
        (g0) -- (g1)
        node[pos=0.85, sloped, below, fill=white, fill opacity=0.75] {$\mathbf{\textcolor{rustorange}{y}}^\top K q = \phi(\underline g, \overline g;\mathbf{\textcolor{rustorange}{y}})$};

    \draw[draw=rustorange, thick, decorate, decoration={brace, amplitude=10pt, raise=2pt}]
        (i1) -- (j1b)
        node[midway, yshift=12pt, sloped, above, fill=white, fill opacity=0.75] {$\Delta(\mathbf{\textcolor{rustorange}{y}})$};

  \end{tikzpicture}
\caption{Support function geometry for DAM and FTR network-feasible polytopes. For each color, the arrows indicate a support normal direction $K^\top y$ in injection space. Supporting hyperplanes (the dashed and dotted lines) are orthogonal to this direction; their points of tangency determine the support values $\phi(\underline f, \overline f;y)$ and $\phi(\underline g, \overline g; y)$; and their signed separation measured along the support normal direction is the alignment gap $\Delta(y)=\phi(\underline{g},\overline{g};y)-\phi(\underline{f},\overline{f};y)$.}
\label{fig:geometric_interpretation}
\end{figure}
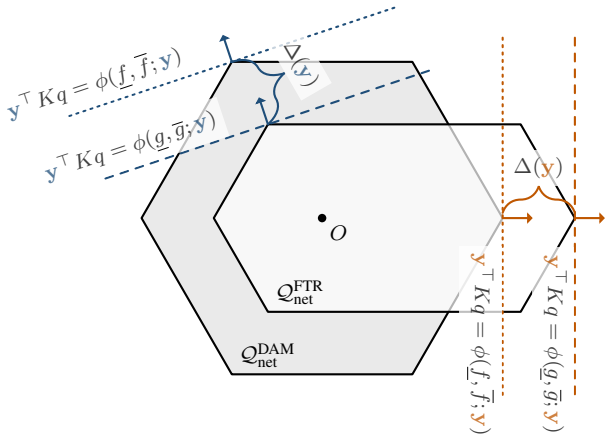

\definition{Alignment Ratio}
When $\phi(\underline f,\overline f;y) > 0$, define the normalized alignment ratio
\begin{equation*}
\eta(y) := \frac{\phi(\underline g,\overline g;y)}{\phi(\underline f,\overline f;y)}
= 1 + \frac{\Delta(y)}{\phi(\underline f,\overline f;y)} .
\end{equation*}

If $\phi(\underline f,\overline f;y)=0$, then $\Delta(y) = 0$ and $\eta(y)$ is undefined.

\proposition{Market Interpretation}\label{prop:support} 
For any dual-optimal DAM shadow-price vector $y^*$,
\begin{align*}
    \mathrm{MS}^{\mathrm{DAM}}(y^*) &= \phi(\underline{f}, \overline{f}; y^*), &&
    \mathrm{PO}^{\mathrm{FTR}}_{\max}(y^*) &= \phi(\underline{g}, \overline{g}; y^*).
\end{align*}

Consequently,
\begin{equation*}
    \Delta(y^*)=\mathrm{PO}^{\mathrm{FTR}}_{\max}(y^*)-\mathrm{MS}^{\mathrm{DAM}}(y^*),
\end{equation*}
so the sign of $\Delta(y^*)$ distinguishes structural underfunding risk ($\Delta(y^*)>0$) from structural hedging inefficiency ($\Delta(y^*)<0$) at realized DAM shadow-price vectors.

Proposition~\ref{prop:support} yields several immediate consequences.
First, alignment is a property of the network models alone: support functions depend only on network topology ($K,b$) and are independent of bids and offers.
Second, alignment is direction-specific: different DAM realizations may induce different DAM shadow-price vectors ($y^*$) and therefore different alignment outcomes under the same FTR network model.
Third, each alignment evaluation reduces to a linear program, enabling tractable ex-post diagnosis.

While Proposition~\ref{prop:support} characterizes alignment at a realized DAM shadow-price vector $y^*$, Definitions~\ref{def:net_support}-\ref{def:alignment} of $\phi(\cdot;\,y)$ and $\Delta(y)$ for arbitrary price vector $y$ enable robust ex ante FTR network model analysis over sets of price vectors.

\section{\label{sec:dual} Dual Attribution and Constraint-Level Decomposition}

This section derives a dual representation of the support-function formulations that attributes alignment gaps to individual transmission element-contingency constraints.

The key observation is that dual feasibility depends only on network geometry $K$ and arbitrary price vector $y$, and not on the limit values $(\underline b,\overline b)$, so DAM and FTR instances share the same dual-feasible set.
Because dual optima are generally non-unique in realistic networks, we later characterize multiplier ranges that are invariant to dual degeneracy.

\subsection{Dual Formulation of the Support Problem}
Recall the network-feasible injection polytope from Definition~\ref{def:network_polytope}, write the two-sided limits as upper bounds, and define the stacked constraint matrix and stacked limit vector
\begin{align*}
    A := \begin{bmatrix} K \\ -K \end{bmatrix} \in \mathbb{R}^{2m\ell\times n},
    \quad
    b := \begin{bmatrix} \overline b \\ -\underline b \end{bmatrix}
    \in (\mathbb{R}\cup\{+\infty\})^{2m\ell}.
\end{align*}

The network support problem from Definition~\ref{def:net_support} can then be written as
\begin{equation*}
    \begin{alignedat}{3}
    \phi(b; y) := 
        &\max_{q \in \mathbb{R}^n}\;& y^\top K q \\
        &\text{subject to}\;&
            (Aq)_i &\le& b_i\quad & \forall i \in \mathcal{J}(b), \\
            & & \mathbf{1}^\top q & = & 0.\quad &
    \end{alignedat}
\end{equation*}
where $\mathcal J(b) := \{i:b_i < +\infty\}$. For brevity, we write $\phi(b;y)$ for the stacked representation.

Let $\mu\in\mathbb R^{2m\ell}_+$ be multipliers on $Aq\le b$ and let $s\in\mathbb R$ be the multiplier on $\mathbf 1^\top q=0$.
Allowing $b_i = +\infty$ implicitly enforces $\mu_i = 0$ in the dual, so we may write $Aq\leq b$ over all indices without loss of generality.
The Lagrangian is
\begin{align*}
    \mathcal L(q,\mu,s) &= (y^\top K-\mu^\top A-s\,\mathbf 1^\top)q + \mu^\top b,
\end{align*}
and the supremum over $q$ is finite if and only if 
\begin{align*}
    \mu^\top A + s\,\mathbf 1^\top = y^\top K,
\end{align*}
in which case $\sup_q \mathcal L(q,\mu,s)=\mu^\top b$.

\proposition{Strong Duality}\label{prop:strong_duality}
Under Assumption~\ref{ass:nonempty},
\begin{equation*}\label{eq:dual_support}
    \phi(b;y)
    = \min_{\mu\in\mathbb R^{2m\ell}_+,\, s\in\mathbb R}\ \mu^\top b
    \quad\text{s.t.}\quad
    \mu^\top A + s\,\mathbf 1^\top = y^\top K.
\end{equation*}

\subsection{Shared Dual-Feasible Set}

\definition{Dual-Feasible Set}  
For any $y \in \mathbb{R}^{m\ell}$, define
\begin{equation*}
    \mathcal{M}(y) := \left\{ (\mu, s) \in \mathbb{R}^{2m\ell}_+ \times \mathbb{R} \,\middle|\, \mu^\top A + s \mathbf{1}^\top = y^\top K \right\}.
\end{equation*}

A key structural property is that the dual feasible set depends only on the network geometry $K$ and arbitrary price vector $y$, and not on the limit vector $b$.

\corolary{Shared Feasibility System}  
Under the extended-limits convention, the DAM and FTR dual support problems share the same feasible set $\mathcal{M}(y^*)$. Moreover, by Proposition~\ref{prop:strong_duality},
\begin{align*}
    \mathrm{MS}^{\mathrm{DAM}}(y^*) &= \min_{(\mu,s) \in \mathcal{M}(y^*)} \mu^\top {f}, \\
    \mathrm{PO}^{\mathrm{FTR}}_{\max}(y^*) &= \min_{(\mu,s) \in \mathcal{M}(y^*)} \mu^\top {g},
\end{align*}
where 
\begin{align*}
    f := \begin{bmatrix}
        \overline{f}\\ -\underline f
    \end{bmatrix},\;
    g := \begin{bmatrix}
        \overline{g}\\ -\underline g
    \end{bmatrix} \in (\mathbb{R}\cup\{+\infty\})^{2m\ell},
\end{align*}
are the stacked DAM and FTR limit vectors.
Thus alignment gaps arise solely from differences in the objective vectors $f$ and $g$, evaluated over common $\mathcal M(y^*)$.

\subsection{Constraint-Level Attribution via Dual Weights}
Dual variables provide a constraint-level explanation of how the support value $\phi(b;y)$ is achieved in direction $K^\top y$: 
a positive $\mu_i$ indicates that constraint $i$ restricts congestion rents in that direction, while $\mu_i=0$ indicates an insensitivity to that limit.

\definition{Optimal Dual Sets}  
For limit vector $b \in \{{f}, {g}\}$ and arbitrary price vector $y$, define
\begin{equation*}
    \mathcal{S}(b; y) := \arg\min_{(\mu,s) \in \mathcal{M}(y)} \mu^\top b.
\end{equation*}

This set is a (possibly higher-dimensional) face of $\mathcal M(y)$.

\proposition{Sensitivity Bounds}\label{prop:sensitivity_bounds} 
For any $(\mu^*, s^*) \in \mathcal{S}(b; y)$ and $\delta > 0$,
\begin{equation*}
    \phi(b \pm \delta e_j; y) \leq \phi(b;y) \pm \delta \mu^*_j,
\end{equation*}
with equality when loosening ($+$) if $\mu^*_j = 0$, and strict inequality when tightening ($-$) if $\mu^*_j > 0$.

Thus any vector of dual weights characterizes local one-sided sensitivity of the support value to tightening or loosening individual constraints.

The dual feasibility condition, $\mu^\top A + s\,\mathbf 1^\top = y^\top K$, requires the support normal $K^\top y$ to lie in the conic hull of the constraint normals (rows of $A\!=\![K;-K]$), up to the power-balance direction $\mathbf 1$. 
In security-constrained network models, this system imposes $n$ equations on $2m\ell+1$ variables; when transmission constraints greatly outnumber nodes ($m\ell \gg n$), the dual-feasible set $\mathcal M(y)$ is high dimensional and optima are typically attained on faces rather than at unique points.
Linear dependencies among PTDF rows — arising from electrically similar contingencies — and repeated constraint limits further expand these faces by introducing directions along which the dual objective $\mu^\top b$ is invariant.
Dual non-uniqueness is therefore structural, motivating a robust attribution framework.

\subsection{Robust Attribution Under Dual Degeneracy}

The dual-optimal set $\mathcal{S}(b; y)$ need not be a singleton, and individual constraint multipliers may vary across optimal dual certificates. 
We therefore characterize constraint importance through robust multiplier ranges, which also offer a two-sided sensitivity of the support value to tightening and loosening the limit of each constraint. 

\definition{Robust Multiplier Bounds}\label{def:robust_bounds} 
For constraint index $i \in \{1, \ldots, 2m\ell\}$, limit vector $b$, and arbitrary price vector $y$, define
\begin{align*}
    \underline{\mu}_{b,i}(y) := \min_{(\mu,s) \in \mathcal{S}(b;y)} \mu_i, &&
    \overline{\mu}_{b,i}(y) := \max_{(\mu,s) \in \mathcal{S}(b;y)} \mu_i.
\end{align*}
which are computable through auxiliary linear programs.

\proposition{Robust Classification}\label{prop:robust}
For each constraint index $i$,
\begin{align*}
    \{\mu_i : (\mu,s)\in\mathcal S(b;y)\} 
    = \left[\underline{\mu}_{b,i}(y),\;\overline{\mu}_{b,i}(y)\right],
\end{align*}
and $i$ falls into exactly one of the following categories:
\begin{enumerate}
    \item \emph{Definitely binding}: $\underline{\mu}_{b,i}(y)>0$.
    \item \emph{Degenerately binding}: $\underline{\mu}_{b,i}(y)=0<\overline{\mu}_{b,i}(y)$.
    \item \emph{Definitely slack}: $\overline{\mu}_{b,i}(y)=0$.
\end{enumerate}

This classification is invariant across all dual optima.
Moreover, the interval endpoints, $(\underline \mu_{b,i}(y),\ \overline \mu_{b,y}(y))$, coincide with the one-sided directional derivatives of $\phi(b;y)$ under loosening and tightening, respectively.

Given a realized DAM outcome with shadow-price vector $y^*$, ISOs can compute robust multiplier bounds $\underline \mu^f_i(y^*), \overline \mu^f_i(y^*)$ and $\underline \mu^g_i(y^*), \overline \mu^g_i(y^*)$ for all transmission constraints using standard LP solvers. 
Constraints with $\underline \mu^f_i(y^*) > 0$ and $\overline \mu^g_i(y^*) = 0$ immediately identify sources of hedging inefficiency, while constraints with $\underline \mu^g_i(y^*) > 0$ and $\overline \mu^f_i(y^*) = 0$ flag underfunding risk contributors. 
This attribution is robust to dual degeneracy and requires no assumptions about PTDF rank or uniqueness.

\section{\label{sec:alignment} Classification of Network Misalignment}

The dual attribution framework yields constraint-level explanations of structural misalignment between DAM and FTR network models. 
We first characterize misalignment induced by a single constraint difference, then apply the result to canonical DAM-FTR model differences: uniform FTR model line derates, extra FTR model contingencies, and unplanned DAM outages.
Finally, we show that multi-interval FTR products induce intrinsic hedging inefficiency even under identical network models.

\definition{Robust Market Dual Ranges}
Fix $y$. Let $\underline\mu^{f}_j(y),\ \overline\mu^{f}_j(y)$ and $\underline\mu^{g}_j(y),\ \overline\mu^{g}_j(y)$ denote the robust dual multiplier bounds from Definition~\ref{def:robust_bounds} for the DAM and FTR models, respectively.

\proposition{Single Constraint Difference}\label{prop:single_constraint}
Suppose $g = f + \delta e_j$ for some $\delta\in\mathbb R$. Then
\begin{align*}
  \sup_{(\mu,s)\in\mathcal S(g;y)} \delta\,\mu_j
  \;\le\;
  \Delta(y)
  \;\le\;
  \inf_{(\mu,s)\in\mathcal S(f;y)} \delta\,\mu_j .
\end{align*}

\corolary{Robust Multiplier Characterization}\label{cor:single_constraint_robust}
If constraint $j$ is looser in the FTR model ($\delta>0$) and $\overline{\mu}_j^g(y)>0$, then $\Delta(y) > 0$, creating underfunding risk precisely when constraint $j$ is definitely or degeneratively binding in the FTR support problem. 

If constraint $j$ is tighter in the FTR model ($\delta<0$) and $\overline{\mu}_j^f(y)>0$, then $\Delta(y) < 0$, creating hedging inefficiency precisely when constraint $j$ is definitely or degeneratively binding in the DAM support problem.

Otherwise, the bounds collapse and $\Delta(y) = 0$.

\subsection{Canonical Network Model Differences}

We accompany this section with a stylized 3-node network model. Details of this toy model are provided in Appendix \ref{app:toy}, along with visualizations of network feasible polytopes (Figures~\ref{fig:toy_feasible}-\ref{fig:toy_align}), and values of $MS^{\text{DAM}}(y^*)$, $PO^{\text{FTR}}_{\max}(y^*)$, $\mu^g(y^*)$, and $\mu^f(y^*)$ for various DAM shadow-price vectors $y^*$ (Tables~\ref{tab:toy_metrics}-\ref{tab:toy_duals}).

A common ISO practice is to apply a uniform derate factor $\alpha\in(0,1]$ to the line limits used in the FTR model \cite{potomac2024ercot}.

\theorem{Uniform Derate}\label{theo:uniform_derate}
If $\mathcal C^{\text{FTR}} = \mathcal C^{\text{DAM}}$ and $g = \alpha f$ for $\alpha\in(0,1]$, then for any DAM shadow-price vector $y^*$ with $\mathrm{MS}^{\mathrm{DAM}}(y^*) > 0$,
\begin{enumerate}
    \item $\mathrm{PO}^{\mathrm{FTR}}_{\max}(y^*) =\alpha\,\mathrm{MS}^{\mathrm{DAM}}(y^*)$,
    \item $\eta(y^*)=\alpha$, and
    \item $\mathcal{S}(g;y^*)=\mathcal{S}(f;y^*)$.
\end{enumerate}

The exact quantification $\eta(y^*) = \alpha$ across all congestion patterns enables precise efficiency-cost tradeoffs. For example, a 25\% uniform derate ($\alpha = 0.75$) reduces hedging capacity to exactly 75\% of available DAM merchandising surplus, providing ISOs with a quantifiable efficiency cost for this conservative decision.

In our toy model, uniformly derating all FTR line limits by $\alpha=0.75$ contracts the feasible region (Figure~\ref{fig:toy_ftr_derate}) and yields $\eta(y^*)=0.75$ across all congestion patterns, with identical DAM and FTR dual multipliers in each pattern (Tables~\ref{tab:toy_metrics}–\ref{tab:toy_duals}).

We next analyze including an additional contingency in the FTR model that is not enforced in the DAM, a conservative modeling choice that can reduce underfunding risk \cite{Hogan2003, Kristiansen2004}.

\theorem{Extra Contingency}\label{theo:extra}
If $c\in\mathcal{C}^{\mathrm{FTR}}\setminus\mathcal{C}^{\mathrm{DAM}}$, then 
\begin{align*}
    \mathcal Q^{\mathrm{FTR}}_{\mathrm{net}}\subseteq\mathcal Q^{\mathrm{DAM}}_{\mathrm{net}},
    \quad\text{and}\quad
    \Delta(y^*)\le0\ \forall y\in\mathbb R^{m\ell},
\end{align*}
with strict inequality ($\Delta(y^*)<0$) if and only if some transmission element constraint $c_i$, under contingency $c$, always satisfies $\mu^g_{c_i} > 0$.

In our toy model, adding an outage contingency to the FTR model (but not the DAM) shrinks the feasible region (Figure~\ref{fig:toy_ftr_extra}) and yields $\Delta(y^*)\le0$ across all congestion patterns (Table~\ref{tab:toy_metrics}). In pattern~(a), the added contingency prices a transmission element in the FTR dual, producing strict hedging inefficiency with $\Delta(y^*)<0$ (Table~\ref{tab:toy_duals}).

Conversely, an unplanned outage in the DAM may result in a contingency being enforced operationally but omitted from the FTR model.

\theorem{Missing Contingency}\label{theo:outage}
If $d\in\mathcal{C}^{\mathrm{DAM}}\setminus\mathcal{C}^{\mathrm{FTR}}$, then
\begin{align*}
    \mathcal Q^{\mathrm{FTR}}_{\mathrm{net}}\supseteq\mathcal Q^{\mathrm{DAM}}_{\mathrm{net}},
    \quad\text{and}\quad
    \Delta(y^*)\ge0\ \forall y\in\mathbb R^{m\ell},
\end{align*}
with strict inequality ($\Delta(y^*)>0$) if and only if some transmission element constraint, $d_j$, under contingency $d$, always satisfies $\mu^f_{d_j} > 0$.

Following an underfunding event, ISOs can identify the responsible outages by checking whether unplanned DAM-only contingency constraints satisfy $\underline{\mu}^f_i(y^*)>0$.
Such constraints restrict DAM merchandising surplus and permit underfunding, providing a diagnostic for improving future FTR outage coordination.

In our toy model, enforcing an additional contingency in the DAM model (but not in the FTR model) shrinks the feasible region (Figure~\ref{fig:toy_dam_outage}) and yields $\Delta(y^*)\ge0$ across all congestion patterns (Table~\ref{tab:toy_metrics}). Underfunding occurs in patterns~(a)–(b), where the missing contingency is priced in every DAM dual optimal; pattern~(c), with no missing contingency constraint priced, yields exact alignment (Table~\ref{tab:toy_duals}).

Missing FTR contingencies can permit underfunding, but never increase hedging inefficiency, while extra contingencies can cause hedging inefficiency, but never increase underfunding risk.
The robust dual conditions above identify precisely when these effects are economically operative for a given DAM shadow-price vector $y^*$.

\subsection{Multi-Interval Temporal Inconsistency}

Consider a multi-interval FTR product spanning periods $\mathcal T=\{1,\dots,T\}$ with realized DAM shadow-price vectors $\{y^*_t\}_{t\in\mathcal T}$, where the FTR auction enforces a \emph{single} injection profile $q$ settled against all intervals.

\theorem{Multi-Interval}\label{theo:multi_int}
If $\mathcal Q^{\mathrm{FTR}}_{\mathrm{net}}=\mathcal Q^{\mathrm{DAM}}_{\mathrm{net}}$, then for any DAM shadow-price vector $y^*$,
\begin{align*}
    \mathrm{PO}^{\mathrm{FTR}}_{\max}\left(\sum_{t\in\mathcal T}y^*_t\right) := \max_{q\in\mathcal Q_{\mathrm{net}}^{\text{FTR}}}\sum_{t\in\mathcal T} y_t^{*\top} K q
    \leq \sum_{t\in\mathcal T} MS^\text{DAM}(y^*_t),
\end{align*}
with equality if and only if all $y^*_t$ support the same face of $\mathcal Q_{\mathrm{net}}^\text{FTR}$.

\corolary{Multi-Interval Alignment Ratio}\label{cor:multi_int_ratio}
The alignment ratio satisfies
\begin{align*}
    \eta_\text{multi}\left(\{y^*_t\}_{t\in\mathcal T}\right) := 
    \frac{
        \mathrm{PO}^{\mathrm{FTR}}_{\max}\left(\sum_{t\in\mathcal T}y^*_t\right)
        }{
        \sum_{t\in \mathcal T} \mathrm{MS}^{\mathrm{DAM}}(y^*_t)
    }  \le 1,
\end{align*}
with strict inequality whenever congestion patterns vary temporally.

Multiple sources of misalignment can be combined cleanly using support-function identities. For example, under a uniform derate $ g=\alpha f$ with no other network differences, Theorem~\ref{theo:uniform_derate} gives
\begin{align*}
    \mathrm{PO}^{\mathrm{FTR}}_{\max}(y^*) = \alpha\,\mathrm{MS}^{\mathrm{DAM}}(y^*),
\end{align*}
and together with Proposition~\ref{prop:support} and Corollary~\ref{cor:multi_int_ratio} we have
\begin{align*}
    \eta_\text{multi}\left(\{y^*_t\}_{t\in\mathcal T}\right)
    = \frac{\alpha\; \phi\left(f;\sum_{t\in\mathcal T} y^*_t\right)}{\sum_t \phi(f;y^*_t)}
    \leq \alpha.
\end{align*}

Additional contingency mismatches can be incorporated via containment bounds (Theorems~\ref{theo:extra}-\ref{theo:outage}) in the same manner; the tight constants depend on which contingencies are priced in the relevant DAM shadow-price vectors.

\section{\label{sec:conclusion} Conclusion}
FTR–DAM misalignment remains a persistent source of hedging inefficiency and underfunding risk in wholesale electricity markets \cite{spp2024som, potomac2024ercot}.
We show that alignment can be characterized geometrically through support functions of network-feasible polytopes, with dual formulations yielding a constraint-level attribution that is robust to dual degeneracy and comparable across market network models.
We apply this framework to canonical FTR-DAM network model differences and to temporal aggregation in FTR contract periods.

Several extensions follow naturally.
The dual-based decomposition can be applied directly to realized market outcomes to diagnose which network constraints drive observed underfunding or hedging inefficiency.
More broadly, the framework enables ex ante FTR network model design by treating contingencies and transmission element limits as tunable parameters that trade off hedging efficiency and revenue adequacy.
Geometrically, alignment is closely related to polyhedral containment, suggesting connections to broader linear-programming theory \cite{Freund1985, sadraddini2019}.
Extending the analysis to richer network representations and to additional inefficiencies arising from strategic FTR bidding remain important directions for future work as well.

\section{AI Usage Disclosure}
All core concepts, theory, and intuition were developed by the authors, as was all final wording in this copy. AI tools were used as an interactive reviewer to clarify technical arguments and proof logic, and to provide feedback on the paper structure, clarity, and presentation, including suggested rewrites of sentences and paragraphs. 

\appendix 

\subsection{\label{app:proofs}Proofs}
\textbf{Proposition~\ref{prop:support} (Support Function Characterization).}
We prove $\mathrm{MS}^{\mathrm{DAM}}(y^*)=\phi(\underline f,\overline f;y^*)$; the FTR identity follows immediately from Definitions~\ref{def:max_ftr_po} and \ref{def:net_support}.

Let $q^{\mathrm{DAM}*}$ be an optimal DAM injection and let $y^{+*},y^{-*}$ be the optimal multipliers on the upper and lower monitored element flow constraints in $\mathcal Q^{\mathrm{DAM}}_{\mathrm{net}}$, with $y^*:=y^{+*}-y^{-*}$.

For any $q\in\mathcal Q^{\mathrm{DAM}}_{\mathrm{net}}$, feasibility gives
$(Kq)_i\le \overline f_i$ for $i\in\mathcal J^+(\overline f)$ and $(Kq)_i\ge \underline f_i$ for $i\in\mathcal J^-(\underline f)$.
Multiplying by $y^{+*}\ge 0$ and $-y^{-*}\le 0$ and summing yields
\begin{align*}
    y^{*\top} K q \le y^{+*\top} \overline{f} - y^{-*\top} \underline{f} .
\end{align*}

By complementary slackness, equality holds at the realized injection $q^{(\mathrm{DAM})*}$, hence
\begin{align*}
y^{*\top} K q^{(\mathrm{DAM})*}
= y^{+*\top} \overline{f} - y^{-*\top} \underline{f}.
\end{align*}

Substituting into the inequality, we obtain
\begin{align*}
    y^{*\top} K q \le y^{*\top} K q^{(\mathrm{DAM})*}
    \quad \forall q \in \mathcal{Q}^{\mathrm{DAM}}_{\mathrm{net}}.
\end{align*}
Therefore $q^{(\mathrm{DAM})*}$ attains the maximum of $y^{*\top} K q$ over $\mathcal{Q}^{\mathrm{DAM}}_{\mathrm{net}}$, so
\begin{align*}
    \mathrm{MS}^{\mathrm{DAM}}(y^*) &= y^{*\top} K q^{(\mathrm{DAM})*}\\
    &= \max_{q \in \mathcal{Q}^{\mathrm{DAM}}_{\mathrm{net}}} y^{*\top} K q  = \phi(\underline{f}, \overline{f}; y^*).
\end{align*}
\hfill$\square$

\textbf{Proposition~\ref{prop:strong_duality} (Strong Duality).}
By Assumption~\ref{ass:nonempty}, the primal feasible set $\mathcal{Q}_{\mathrm{net}}(\underline{b},\overline{b})$ is nonempty and compact. The objective is linear, hence the primal optimum is finite and attained. Strong duality and dual attainment then follow from standard linear programming duality theorems for feasible and bounded LPs \cite{bertsimas1997}.
\hfill$\square$

\textbf{Proposition~\ref{prop:sensitivity_bounds} (Sensitivity Bounds).}
Fix $y$ and $b$, and let $(\mu^*,s^*)\in\mathcal S(b;y)$.
For any perturbation $b'$, the feasible set $\mathcal M(y)$ is unchanged because it depends only on $(K,y)$ and not on $b$.
Hence $(\mu^*,s^*)\in\mathcal M(y)$ is feasible for the dual problem with objective $\mu^\top b'$.
By optimality,
\begin{align*}
    \phi(b';y)=\min_{(\mu,s)\in\mathcal M(y)} \mu^\top b' \;\le\; \mu^{*\top} b'.
\end{align*}

For tightening with $b' = b-\delta e_j$ ($\delta>0$),
\begin{align*}
    \phi(b-\delta e_j;y)\le \mu^{*\top}(b-\delta e_j)
    = \mu^{*\top}b-\delta \mu^*_j
    = \phi(b;y)-\delta\mu^*_j.
\end{align*}

If $\mu^*_j>0$, the right-hand side is strictly less than $\phi(b;y)$.

For loosening with $b' = b+\delta e_j$ ($\delta>0$),
\begin{align*}
    \phi(b+\delta e_j;y)\le \mu^{*\top}(b+\delta e_j)
    = \phi(b;y)+\delta\mu^*_j.
\end{align*}

If $\mu^*_j=0$, this gives $\phi(b+\delta e_j;y)\le \phi(b;y)$.
Since the primal feasible set expands when $b$ is loosened, we also have $\phi(b+\delta e_j;y)\ge \phi(b;y)$, hence equality holds.
\hfill$\square$

\textbf{Proposition~\ref{prop:robust} (Robust Classification).}
The set $\mathcal S(b;y)$ is convex, and projection onto coordinate $i$ preserves convexity; hence the image
\begin{align*}
    \{\mu_i : (\mu,s)\in\mathcal S(b;y)\},
\end{align*}
is an interval. Because $\mathcal S(b;y)$ is a closed face of a polyhedron, the extrema $\underline{\mu}_{b,i}(y)$ and $\overline{\mu}_{b,i}(y)$ are attained, so the interval is closed.
Moreover, the support function $\phi(b;y)$ is convex in $b$, and its subdifferential in index $i$ satisfies
\begin{align*}
    \partial_{b_i}\phi(b;y)
    = \{\mu_i : (\mu,s)\in\mathcal S(b;y)\},
\end{align*}
so the directional derivative with respect to constraint index $i$ is generally set-valued, allowing for multiple dual-optimal explanations of the same support value.

With this proved, the classification follows immediately:
1) $\underline\mu_{b,i}(y)>0$ iff $\mu_i>0$ in every optimal dual certificate (definitely binding);
2) $\underline\mu_{b,i}(y)=0<\overline\mu_{b,i}(y)$ iff some optimal duals assign $\mu_i=0$ and others assign $\mu_i>0$ (degenerately binding);
3) $\overline\mu_{b,i}(y)=0$ iff $\mu_i=0$ in every optimal dual certificate (definitely slack).

Finally, since $\phi(b;y)$ is convex in $b$ and $\mathcal S(b;y)$ is its subdifferential, the one-sided directional derivatives in directions $\pm e_i$ equal the extremal subgradients, i.e.,
$\underline\mu_{b,i}(y)$ (loosening) and $\overline\mu_{b,i}(y)$ (tightening).
\hfill$\square$

\textbf{Proposition~\ref{prop:single_constraint} (Single Constraint Difference).}
Fix $y$ and suppose $g=f+\delta e_j$.
Since $\mathcal S(f;y),\mathcal S(g;y)\subseteq \mathcal M(y)$, any $(\mu^f,s^f)\in\mathcal S(f;y)$ is feasible for the $g$-dual problem, and any $(\mu^g,s^g)\in\mathcal S(g;y)$ is feasible for the $f$-dual problem. Hence
\begin{align*}
    \phi(g;y) &\leq (\mu^f)^\top g = (\mu^f)^\top f + \delta\,\mu^f_j = \phi(f;y) + \delta\,\mu^f_j,\\
    \phi(f;y) &\leq (\mu^g)^\top f = (\mu^g)^\top g - \delta\,\mu^g_j = \phi(g;y) - \delta\,\mu^g_j.
\end{align*}
Rearranging gives $\delta\,\mu^g_j \le \Delta(y) \le \delta\,\mu^f_j$ for all choices $(\mu^f,s^f)\in\mathcal S(f;y)$ and $(\mu^g,s^g)\in\mathcal S(g;y)$.
Taking $\sup$ over $\mathcal S(g;y)$ on the left and $\inf$ over $\mathcal S(f;y)$ on the right yields the claim. 

In Corollary~\ref{cor:single_constraint_robust}, the active robust multiplier is determined by the sign of $\delta$.
If $\delta>0$, then
\begin{align*}
    \max_{(\mu,s)\in\mathcal S(g;y)} \delta\,\mu_j = \delta\,\overline{\mu}_j^g(y),
    \qquad
    \min_{(\mu,s)\in\mathcal S(f;y)} \delta\,\mu_j = \delta\,\underline{\mu}_j^f(y),
\end{align*}
while if $\delta<0$ the extrema select the opposite endpoints.
By monotonicity, $\Delta(y)$ has the same sign as $\delta$; thus, if the corresponding upper robust bound is zero, both bounds equal zero and $\Delta(y)=0$.
\hfill$\square$

\textbf{Theorem~\ref{theo:uniform_derate} (Uniform Derate).}
By Corollary~1, both the DAM and FTR support problems minimize linear objectives over the same dual-feasible set $\mathcal M(y^*)$.
Since $g=\alpha f$ with $\alpha>0$, we have
\begin{align*}
    \arg\min_{(\mu,s)\in\mathcal M(y^*)} \mu^\top g
    &= \arg\min_{(\mu,s)\in\mathcal M(y^*)} \alpha\,\mu^\top f \\
    &= \arg\min_{(\mu,s)\in\mathcal M(y^*)} \mu^\top f,
\end{align*}
and therefore 3) $\mathcal S(g;y^*)=\mathcal S(f;y^*)$.
Moreover, for any $(\mu,s)\in\mathcal M(y^*)$, $\mu^\top g = \alpha\,\mu^\top f$, so evaluating at an optimizer yields 1)
\begin{align*}
    \mathrm{PO}^{\mathrm{FTR}}_{\max}(y^*)
    = \phi(g;y^*)
    = \alpha\,\phi(f;y^*)
    = \alpha\,\mathrm{MS}^{\mathrm{DAM}}(y^*).
\end{align*}

The expression for 2) $\eta(y^*) = \alpha$ follows immediately.
\hfill$\square$

\textbf{Theorem~\ref{theo:extra} (Extra Contingency).}
Enforcing an additional contingency adds transmission constraints, so $\mathcal Q^{\mathrm{FTR}}_{\mathrm{net}}\subseteq\mathcal Q^{\mathrm{DAM}}_{\mathrm{net}}$.
By monotonicity of support functions under set inclusion, $\phi(g;y)\le \phi(f;y)$ for all $y$, hence $\Delta(y^*)\le 0$.

Suppose there exists a dual optimal $(\mu,s)\in\mathcal S(g;y^*)$ such that $\mu_{c_i} = 0$ for every $c_i$ under contingency $c$.
Proposition~\ref{prop:sensitivity_bounds} implies that loosening all $c$-constraints does not change $\phi(g;y^*)$; since the DAM model omits $c$ (equivalently loosening all $c_i$ to infinite bounds), this yields $\phi(g;y^*)=\phi(f;y^*)$ and hence $\Delta(y^*)=0$.

Conversely, suppose that for every dual optimal $(\mu,s)\in\mathcal S(g;y^*)$, there exists at least one constraint $c_i$ under contingency $c$ with $\mu_{c_i}>0$. 
If equality $\phi(g;y^*)=\phi(f;y^*)$ were to hold, then by Proposition~\ref{prop:sensitivity_bounds} there would exist a dual optimal with $\mu_{c_i}=0$ for all $c_i$, contradicting the assumption. Therefore $\phi(g;y^*)<\phi(f;y^*)$ and  $\Delta(y^*)<0$.
\hfill$\square$

\textbf{Theorem~\ref{theo:outage} (Missing Contingency).}
The argument mirrors the proof of Theorem~\ref{theo:extra}. Omitting contingency $d$ removes transmission constraints, so $\mathcal Q^{\mathrm{FTR}}_{\mathrm{net}}\supseteq\mathcal Q^{\mathrm{DAM}}_{\mathrm{net}}$ and $\Delta(y^*)\ge 0$.

If there exists a constraint under $d$ that is priced in every DAM-optimal dual, this yields $\Delta(y^*)>0$ since the FTR model omits $d$ (equivalent to loosening all DAM $d_j$ to infinite bounds). Conversely, if no such constraints exist, then there is a DAM-optimal dual assigning zero weight to all constraints under $d$, so omitting $d$ does not change the support value and $\Delta(y^*) = 0$.
\hfill$\square$

\textbf{Theorem~\ref{theo:multi_int} (Multi-Interval Inefficiency).}
By definition,
\begin{align*}
    \mathrm{PO}^{\mathrm{FTR}}_{\max}\!\left(\sum_{t\in\mathcal T} y_t^*\right)
    = \max_{q\in\mathcal Q_{\mathrm{net}}} \left(\sum_{t\in\mathcal T} y_t^*\right)^\top K q
    = \phi\!\left(b;\sum_{t\in\mathcal T} y^*_t\right).
\end{align*}

Since $\phi(\cdot)$ is a support function, it is subadditive, so
\begin{align*}
    \phi\!\left(b;\sum_{t\in\mathcal T} y^*_t\right)
    \le \sum_{t\in\mathcal T} \phi(b;y^*_t)
    = \sum_{t\in\mathcal T} \mathrm{MS}^{\mathrm{DAM}}(y^*_t).
\end{align*}

Equality holds if and only if there exists a single injection $q^*\in\mathcal Q_{\mathrm{net}}^\text{FTR}$ that simultaneously attains $\phi(b;y^*_t)$ for all $t\in\mathcal T$, i.e., all $y^*_t$ support the same face of $\mathcal Q_{\mathrm{net}}^\text{FTR}$.
\hfill$\square$

\subsection{\label{app:toy} Toy Network Model}
We use the stylized three-node network model in Figure~\ref{fig:toy_model} to
illustrate the canonical sources of DAM–FTR misalignment analyzed in Section~\ref{sec:alignment}.
This appendix defines the toy model, constructs the associated network-feasible sets, and documents the congestion patterns used to evaluate alignment and dual attribution.
All interpretation is deferred to the main text.

\begin{figure}[h]
    \centering
    {\footnotesize
    \begin{tikzpicture}[
    scale=0.7, line join=round, line cap=round,
  bus/.style={circle, draw, thick, minimum size=8mm, inner sep=0pt}
]
  \coordinate (L) at (3, 0);
  \coordinate (S) at (0, 3);
  \coordinate (C) at (6, 3); 

\node[bus,label=center:$S$] (nS) at (S) {};
  \node[bus,label=center:$C$] (nC) at (C) {};
  \node[bus,label=center:$L$] (nL) at (L) {};

  \draw[thick] (nS) -- (nL) node[sloped, above, midway] {$|f_{SL}|\leq75\text{MW}$};
  \draw[thick] (nC) -- (nL);
  \draw[thick] (nS) -- (nC) node[midway,below] {$|f_{SC}|\leq25\text{MW}$};

  \draw[->,thick,shorten >=1pt] (nL.west) -- (2, 0)
    node[near end,left] {$\mathcolor{blue}{\mathbf{q_L}}$};
  \draw[->,thick,shorten >=1pt] ($(nS.north) + (0, 0.5)$) -- (nS.north)
    node[near start,above,align=center, text width=3cm] {$p_S= \$5/\text{MWh}$\\ $q_S \in [0,\, \mathcolor{blue}{\mathbf{\overline{q_S}}}]\text{MW}$};
  \draw[->,thick,shorten >=1pt] ($(nC.north) + (0, 0.5)$) -- (nC.north)
    node[near start,above,align=center, text width=3cm] {$p_C=\$150/\text{MWh}$\\ $q_C \in [0,\, 300]\text{MW}$};
\end{tikzpicture}
    }
    \caption{Three-node toy model with nodes $S$ (solar), $C$ (coal), and $L$ (load), connected in a triangle. Line reactances are normalized to one. Lines $SL$ and $SC$ have finite thermal limits, while line $CL$ is unconstrained.}
    \label{fig:toy_model}
\end{figure}
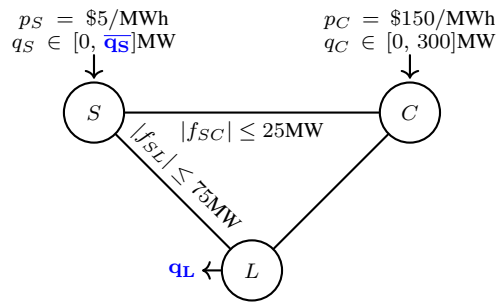

For a given load $q_L \in \mathbb R_+$, feasible dispatch is described by injections
$q_S,q_C \in \mathbb R_+$ satisfying
\begin{align*}
  q_S &+ q_C = q_L, && \text{(power balance)}\\
  q_S &\le \overline{q_S}, \quad q_C \le 300, && \text{(generator limits)}\\
  |f_{SL}| &= \bigl|\tfrac{2}{3} q_S + \tfrac{1}{3} q_C\bigr| \le 75, && \text{(line $SL$ limits)}\\
  |f_{SC}| &= \bigl|\tfrac{1}{3} q_S - \tfrac{1}{3} q_C\bigr| \le 25, && \text{(line $SC$ limits)}
\end{align*}
where line flows follow from DC power flow with unit reactances.
Coal capacity $\overline q_C = 300$ is chosen sufficiently large that it does not bind in the scenarios considered, while solar capacity $\overline q_S$ is varied to induce congestion.

Eliminating $q_C = q_L - q_S$ yields the equivalent constraints
\begin{align*}
  q_S + q_L &\le 225,\\
  -75 \le 2q_S - q_L &\le 75,
\end{align*}
together with $0 \le q_S \le \overline{q_S}$ and $q_L \ge 0$.
These inequalities define the feasible region in
$(q_L,q_S)$-space shown in Figure~\ref{fig:toy_feasible}.

\begin{figure}[h!]
    \centering
    \includegraphics[width=1.0\linewidth]{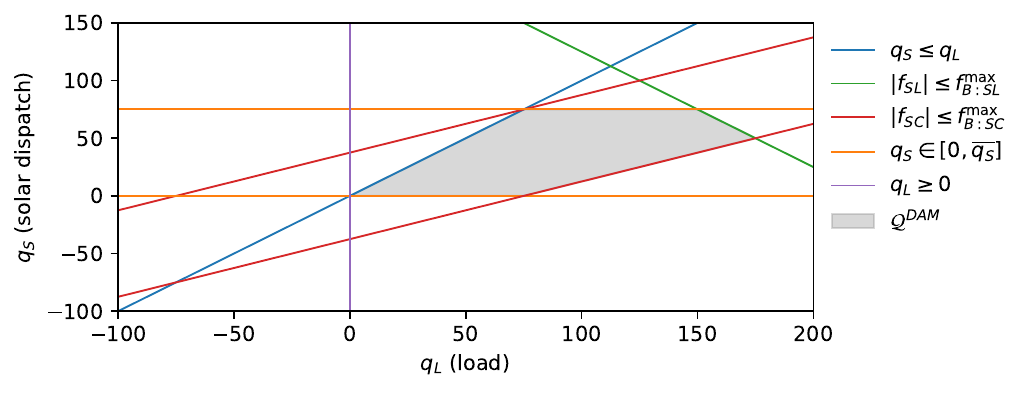}
    \caption{Network-feasible load $q_L$ vs solar dispatch $q_S$ for the base network with $\overline{q_S}=75$~MW.}
    \label{fig:toy_feasible}
\end{figure}

We construct three network model difference variations corresponding to the misalignment mechanisms studied in Section~\ref{sec:alignment}:
\begin{itemize}
    \item \emph{Uniform FTR derate:} DAM and FTR share the base network, but all FTR line limits are uniformly reduced to $75\%$ of DAM limits.
    \item \emph{Extra FTR contingency:} the FTR model includes an additional outage of line $SL$, while the DAM model enforces only the base network.
    \item \emph{Unplanned DAM outage:} the DAM model includes an additional outage of line $SC$, while the FTR model enforces only the base network.
\end{itemize}

Network-feasible injection polytopes for each difference are shown in Figure~\ref{fig:toy_align}.

\begin{figure}[h!]
    \centering

    \begin{subfigure}{1.0\linewidth}
        \centering
        \includegraphics[width=\linewidth]{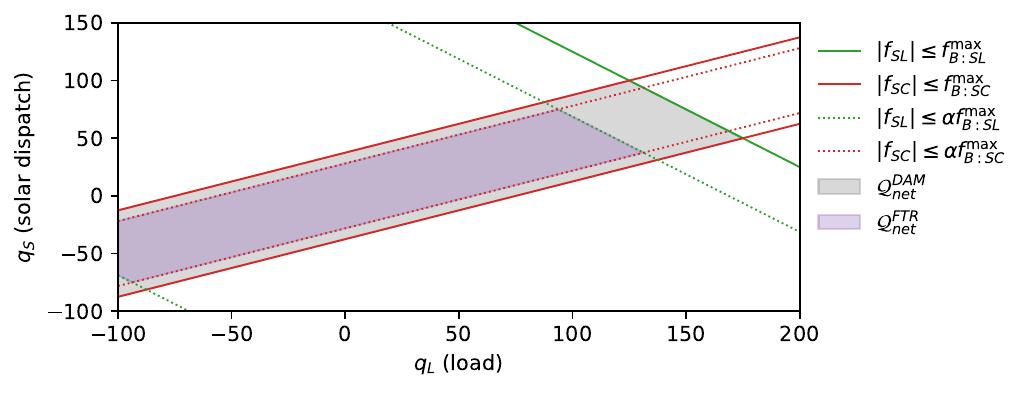}
        \caption{Uniform FTR Derate}
        \label{fig:toy_ftr_derate}
    \end{subfigure}

    \vspace{0.5em}
    \begin{subfigure}{1.0\linewidth}
        \centering
        \includegraphics[width=\linewidth]{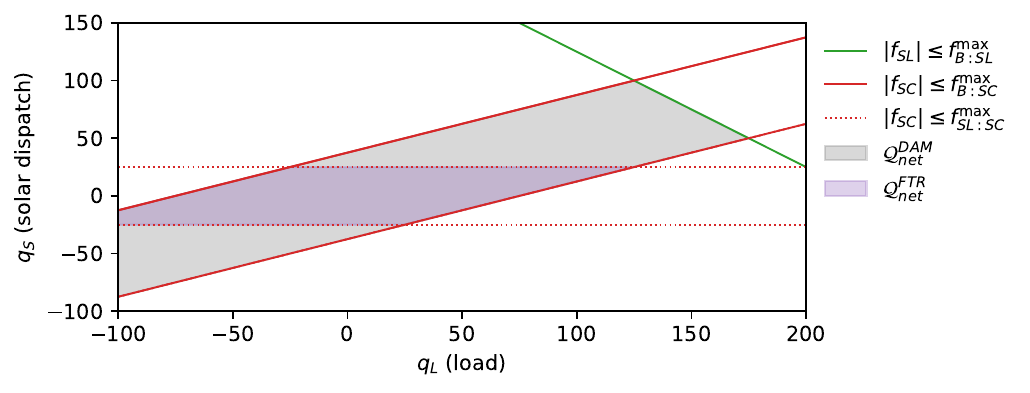}
        \caption{Extra FTR Contingency}
        \label{fig:toy_ftr_extra}
    \end{subfigure}

    \vspace{0.5em}
    \begin{subfigure}{1.0\linewidth}
        \centering
        \includegraphics[width=\linewidth]{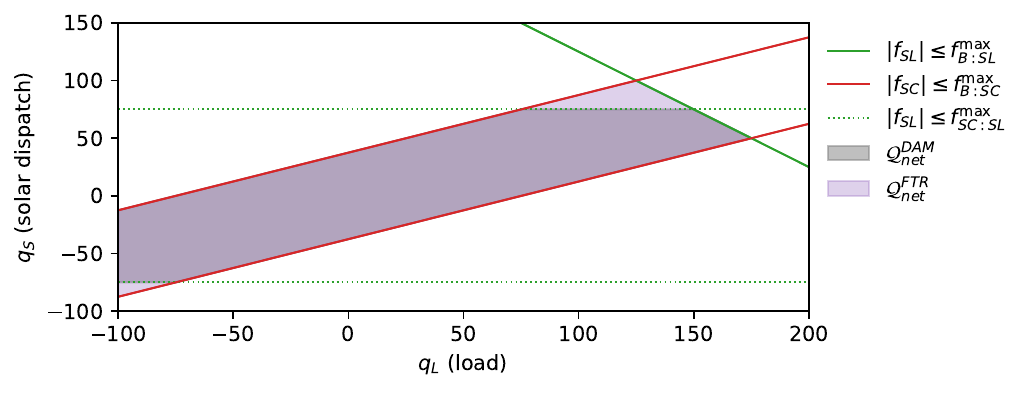}
        \caption{Unplanned DAM Outage}
        \label{fig:toy_dam_outage}
    \end{subfigure}

    \caption{Network-feasible injection polytopes under alternative DAM–FTR modeling assumptions. Solid boundaries indicate constraints enforced in the base case; dotted boundaries indicate constraints enforced under an outage.}
    \label{fig:toy_align}
\end{figure}

Support functions are evaluated for DAM shadow-price vectors $y^*$ corresponding to the congestion patterns in Table~\ref{tab:toy_patterns}.

\begin{table}[h!]
    \caption{Congestion patterns. Entries such as $+\!SL$ and $-\!SC$ indicate binding upper or lower limits on the corresponding line under the indicated contingency.}
    \centering
    \begin{tabular}{c|c|c|c}
    \hline
    \textbf{Pattern} 
        & \textbf{Base ($B\!:$)} 
        & \textbf{SL outage ($SL\!:$)} 
        & \textbf{SC outage ($SC\!:$)} \\
    \hline
    (a) & $+SL$ & \textcolor{gray}{--}  & \textcolor{gray}{--} \\
    (b) & $+SC$ & \textcolor{gray}{--} & \textcolor{gray}{--} \\
    (c) & $-SC$ & \textcolor{gray}{--} & \textcolor{gray}{--} \\
    \hline
    \end{tabular}
    \label{tab:toy_patterns}
\end{table}

Tables~\ref{tab:toy_metrics} and~\ref{tab:toy_duals} report alignment gaps and support-function dual values across all toy model differences and congestion patterns.

\begin{table}[h!]
    \caption{FTR-DAM alignment by model difference and congestion pattern.}
    \label{tab:toy_metrics}
    \centering
    \begin{tabular}{llrrrr}
\toprule
 &  & $MS^{DAM}(y^*)$ & $\Delta(y^*)$ & $\eta(y^*)$ \\
\midrule
\multirow[t]{3}{*}{FTR Derate} & (a) & 32,625 & \textcolor{red}{(8,156)} & 0.75 \\
 & (b) & 5,438 & \textcolor{red}{(1,359)} & 0.75 \\
 & (c) & 1,484 & \textcolor{red}{(371)} & 0.75 \\
\cmidrule(lr){1-5}
\multirow[t]{3}{*}{Extra FTR Cont.} & (a) & 32,625 & \textcolor{red}{(10,875)} & 0.67 \\
 & (b) & 5,438 & 0 & 1.00 \\
 & (c) & 1,484 & 0 & 1.00 \\
\cmidrule(lr){1-5}
\multirow[t]{3}{*}{DAM Outage} & (a) & 16,583 & 2,674 & 1.16 \\
 & (b) & 10,875 & 3,625 & 1.33 \\
 & (c) & 1,101 & 0 & 1.00 \\
\bottomrule
\end{tabular}

\end{table}

\begin{table}[h!]
    \caption{Support-function dual values by model difference and congestion pattern.}
    \label{tab:toy_duals}
    \centering
    \begin{tabular}{P{2.8em} P{0.3em} P{0.6em} | >{\centering\arraybackslash}p{2.2em} >{\centering\arraybackslash}p{2.2em} >{\centering\arraybackslash}p{2.2em} >{\centering\arraybackslash}p{2.2em} >{\centering\arraybackslash}p{2.2em} >{\centering\arraybackslash}p{2.2em}}
\toprule
 &  &  & B:SL & B:SC & SL:SL & SL:SC & SC:SL & SC:SC \\
\midrule
\multirow[t]{6}{*}{\parbox[t]{\linewidth}{FTR Derate}} & \multirow[t]{2}{*}{\parbox[t]{\linewidth}{(a)}} & $\mu^{f}$ & 435 & 0 & \textcolor{gray}{-} & \textcolor{gray}{-} & \textcolor{gray}{-} & \textcolor{gray}{-} \\
 &  & $\mu^{g}$ & 435 & 0 & \textcolor{gray}{-} & \textcolor{gray}{-} & \textcolor{gray}{-} & \textcolor{gray}{-} \\
\cmidrule(lr){2-9}
 & \multirow[t]{2}{*}{\parbox[t]{\linewidth}{(b)}} & $\mu^{f}$ & 0 & 217 & \textcolor{gray}{-} & \textcolor{gray}{-} & \textcolor{gray}{-} & \textcolor{gray}{-} \\
 &  & $\mu^{g}$ & 0 & 217 & \textcolor{gray}{-} & \textcolor{gray}{-} & \textcolor{gray}{-} & \textcolor{gray}{-} \\
\cmidrule(lr){2-9}
 & \multirow[t]{2}{*}{\parbox[t]{\linewidth}{(c)}} & $\mu^{f}$ & 0 & \textcolor{red}{(59)} & \textcolor{gray}{-} & \textcolor{gray}{-} & \textcolor{gray}{-} & \textcolor{gray}{-} \\
 &  & $\mu^{g}$ & 0 & \textcolor{red}{(59)} & \textcolor{gray}{-} & \textcolor{gray}{-} & \textcolor{gray}{-} & \textcolor{gray}{-} \\
\cmidrule(lr){1-9}
\multirow[t]{6}{*}{\parbox[t]{\linewidth}{Extra FTR Cont.}} & \multirow[t]{2}{*}{\parbox[t]{\linewidth}{(a)}} & $\mu^{f}$ & 435 & 0 & 0 & 0 & \textcolor{gray}{-} & \textcolor{gray}{-} \\
 &  & $\mu^{g}$ & 0 & \textcolor{red}{(435)} & 0 & 435 & \textcolor{gray}{-} & \textcolor{gray}{-} \\
\cmidrule(lr){2-9}
 & \multirow[t]{2}{*}{\parbox[t]{\linewidth}{(b)}} & $\mu^{f}$ & 0 & 218 & 0 & 0 & \textcolor{gray}{-} & \textcolor{gray}{-} \\
 &  & $\mu^{g}$ & 0 & 217 & 0 & 0 & \textcolor{gray}{-} & \textcolor{gray}{-} \\
\cmidrule(lr){2-9}
 & \multirow[t]{2}{*}{\parbox[t]{\linewidth}{(c)}} & $\mu^{f}$ & 0 & \textcolor{red}{(59)} & 0 & 0 & \textcolor{gray}{-} & \textcolor{gray}{-} \\
 &  & $\mu^{g}$ & 0 & \textcolor{red}{(59)} & 0 & 0 & \textcolor{gray}{-} & \textcolor{gray}{-} \\
\cmidrule(lr){1-9}
\multirow[t]{6}{*}{\parbox[t]{\linewidth}{DAM Outage}} & \multirow[t]{2}{*}{\parbox[t]{\linewidth}{(a)}} & $\mu^{f}$ & 114 & 0 & \textcolor{gray}{-} & \textcolor{gray}{-} & 107 & 0 \\
 &  & $\mu^{g}$ & 221 & 107 & \textcolor{gray}{-} & \textcolor{gray}{-} & 0 & 0 \\
\cmidrule(lr){2-9}
 & \multirow[t]{2}{*}{\parbox[t]{\linewidth}{(b)}} & $\mu^{f}$ & 0 & 0 & \textcolor{gray}{-} & \textcolor{gray}{-} & 145 & 0 \\
 &  & $\mu^{g}$ & 145 & 145 & \textcolor{gray}{-} & \textcolor{gray}{-} & 0 & 0 \\
\cmidrule(lr){2-9}
 & \multirow[t]{2}{*}{\parbox[t]{\linewidth}{(c)}} & $\mu^{f}$ & 0 & \textcolor{red}{(44)} & \textcolor{gray}{-} & \textcolor{gray}{-} & 0 & 0 \\
 &  & $\mu^{g}$ & 0 & \textcolor{red}{(44)} & \textcolor{gray}{-} & \textcolor{gray}{-} & 0 & 0 \\
\bottomrule
\end{tabular}

\end{table}

In Table~\ref{tab:toy_duals}, reported dual values correspond to the difference between multipliers on the upper and lower limits of each line; line $CL$ is omitted since it has no binding constraint.
Because the feasible injection space is two-dimensional, the support problems admit a unique optimal dual for generic parameter choices, so robust and non-robust dual values coincide in this example.
Although the DAM clearing problem and the DAM support problem have different objectives, the realized DAM shadow-price vector $y^*$ satisfies the support dual optimality conditions in this toy model.
In higher-dimensional networks, multiple dual certificates are typical and exact coincidence need not occur.

\bibliographystyle{ieeetr}
\bibliography{references}
\balance

\endgroup
\end{document}